\documentclass[preprint,aps,floatfix,longbibliography,prapplied]{revtex4-1}

\usepackage[pdftex]{graphicx}% Include figure files
\usepackage{dcolumn}% Align table columns on decimal point
\usepackage{bm}% bold math
\usepackage{color}
\usepackage{xcolor}
\usepackage[shortlabels]{enumitem}

\begin{document}

\title{A way to measure the dispersion forces near the van der Waals-Casimir transition}
%Van der Waals - Casimir forces: A proposal to measure the dispersion forces near the retarded-nonretarded transition
\author{V. B. Svetovoy}
\email[Corresponding author: ]{v.svetovoy@utwente.nl; svetovoy@yandex.ru}
\affiliation{A. N. Frumkin Institute of Physical Chemistry and Electrochemistry, Russian Academy of Sciencies, Leninsky prospect 31 bld. 4, 119071 Moscow, Russia}
\author{A. V. Postnikov}
\affiliation{Valiev Institute of Physics and Technology, Yaroslavl Branch, Russian Academy of Sciences, Universitetskaya 21, 150007 Yaroslavl, Russia}
\author{I. V. Uvarov}
\affiliation{Valiev Institute of Physics and Technology, Yaroslavl Branch, Russian Academy of Sciences, Universitetskaya 21, 150007 Yaroslavl, Russia}
\author{F. I. Stepanov}
\affiliation{Ishlinsky Institute for Problems in Mechanics, Russian Academy of Sciences, prospect Vernadskogo, 101-1, 119526 Moscow, Russia}
\author{G. Palasantzas}
\affiliation{Zernike Institute for Advanced Materials, University of Groningen - Nijenborgh 4, 9747 AG Groningen, The Netherlands}

\begin{abstract}
Forces induced by quantum fluctuations of electromagnetic field control adhesion phenomena between rough solids when the bodies are separated by distances $\sim 10\;$nm. However, this distance range remains largely unexplored experimentally in contrast with the shorter (van der Waals forces) or the longer (Casimir forces) separations. The reason for this is the pull-in instability of the systems with the elastic suspension that poses a formidable limitation. In this paper we propose a genuine experimental configuration that does not suffer from the short distance instability. The method is based on adhered cantilever, whose shape is sensitive to the forces acting near the adhered end. The general principle of the method, its possible realization and feasibility are extensively discussed. The dimensions of the cantilever are determined by the maximum sensitivity to the forces.  If the adhesion is defined by strong capillary or chemical interactions, the method loses its sensitivity. Special discussion is presented for the determination of the minimum distance between the rough solids upon contact, and for the compensation of the residual electrostatic contribution. The proposed method can be applied to any kind of solids (metals, semiconductors, dielectrics) and to any intervening medium (gas or liquid).
\end{abstract}

\maketitle

\section{Introduction}

The interaction between three dimensional bodies induced by quantum fluctuations of the electromagnetic field becomes important at short separations, less than $100\;$nm. This interaction results in attractive dispersion forces (DF) \cite{Casimir1948,Mahanty1976}, which include van der Waals (vdW) and Casimir forces. Lifshitz and co-authors \cite{Lifshitz1956,DLP} demonstrated that both forces are the asymptotics of one and the same force at long (Casimir) and at short (vdW) distances. The DF are actively investigated for the last 20 years \cite{Klimchitskaya2009,Rodriguez2011}.  In a series of critical experiments \cite{Lamoreaux1997,Harris2000,Chan2001,Decca2003,Decca2005} the forces were measured at distances $\gtrsim100\:$nm with a high precision of $\sim1\%$. The results agree with the prediction of the Lifshitz theory \cite{Lifshitz1956,DLP} that accounts for finite temperature and dielectric response of the interacting bodies.  The latter dependence was checked in a number of experiments performed for different materials  \cite{Iannuzzi2004,Chen2007,Man2009,Torricelli2010,Torricelli2012}. The contribution of thermal fluctuations to the force is not important at distances $d\lesssim 100\;$nm.

All the modern experiments were performed with the use of microelectromechanical systems (MEMS), which are characterized by small separation between elements and large areas of these elements. At these conditions (small distance and large area) the DF manifest themselves most distinctly. Irreversible adhesion (stiction) of moving MEMS elements induced by the DF results in failure of MEMS devices \cite{Tas1996,Maboudian1997,Parker2005}. During operation of the device, stiction can occur due to the pull-in instability and at the fabrication stage, it can be induced by capillary forces. The stiction problem seriously restricts applications of MEMS. It worth to note that strong adhesion by chemical interactions or by capillary forces can be excluded by special preparation of the surfaces, but a relatively weak adhesion due to the DF cannot be excluded in principle. The latter adhesion is responsible for many natural phenomena (for example, a firm grip of geckos on walls and ceilings \cite{Autumn2002}) and for many artificial processes (for example, stiction of a polymeric film to surfaces \cite{Bhushan2003}). This adhesion forms an entire class of physical phenomena, which appear often in the surrounding world. The DF produce not only negative effect on MEMS. Actuation of MEMS devices with the vdW/Casimir forces is a dynamically developing field \cite{Ball2007,Capasso2007,Buks2001,Palasantzas2005a,Palasantzas2005b,DelRio2005,Esquivel2006,Esquivel2006,Esquivel2009,Broer2013,Broer2015,Sedighi2015}.

\begin{figure}[ptb]
\begin{center}
%\vspace{0.3cm}
\includegraphics[width=0.7\textwidth]{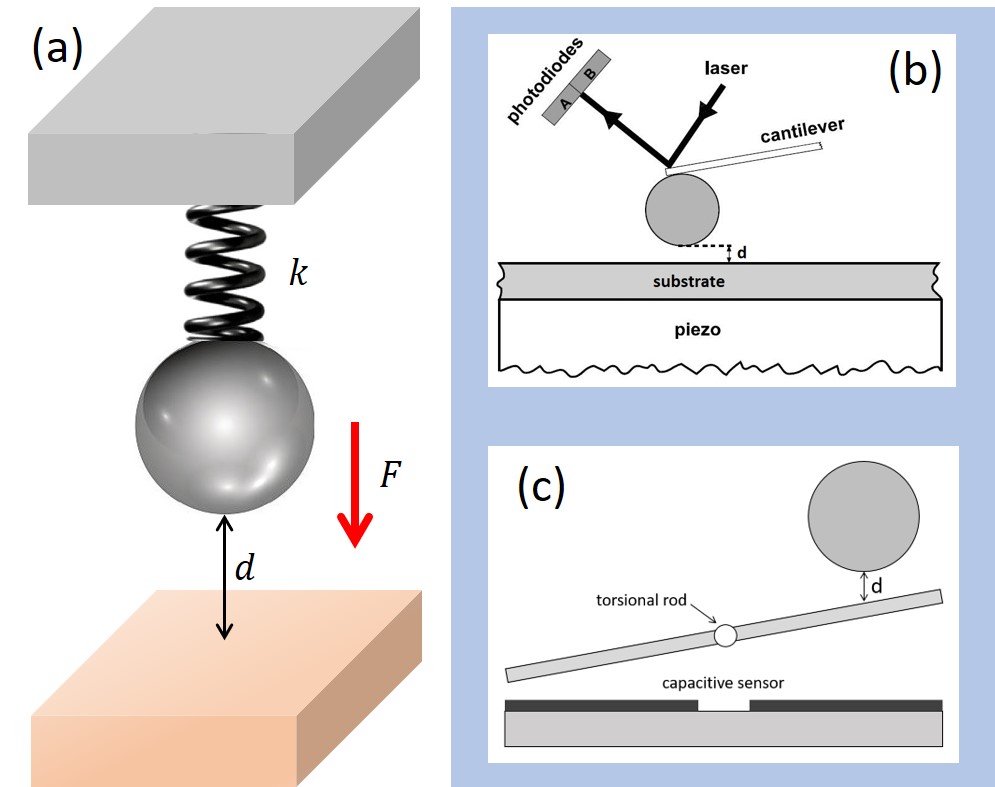}
\caption{(a) Schematic view for the standard way to measure the force between a sphere and a plate. The force $F$ is balanced by a spring of stiffness $k$. (b) AFM realization \cite{Harris2000}. (c) Torsional rod realization \cite{Chan2001}. \label{fig:spring}}
\end{center}
\end{figure}

When two solids get into contact they are still separated by the average distance $d\sim10\;$nm due to finite roughness of contacting bodies. This distance range corresponds to the transition between vdW and Casimir forces that physically means the transition between retarded and unretarded interaction. This range was studied much less than the range $d\gtrsim100\;$nm, although the DF at shorter distances play more important role. Poor knowledge of the forces in the transition region is related to the same pull-in instability that leads to stiction of the MEMS elements. In all the experiments the force is balanced by a spring of a known stiffness $k$ as shown schematically in Fig.~\ref{fig:spring}. A principal disadvantage of such a system is the loss of stability at short separations. In some experiments the spring is a cantilever of the atomic force microscope (AFM) \cite{Harris2000,Chen2007,Torricelli2010,Torricelli2012}. In other experiments this role belongs to a torsional rod \cite{Chan2001,Decca2003,Decca2005,Man2009} or to a string \cite{Lamoreaux1997}. In all experiments but one \cite{Bressi2002} the force was measured between a sphere and plate to avoid the parallelism problem. Due to strong surface charging of dielectrics the force was measured between well conductive materials or doped semiconductors \cite{Lamoreaux1997,Harris2000,Chan2001,Decca2003,Decca2005,Iannuzzi2004,Chen2007,Man2009,Torricelli2010}.

In practice the shortest jump-to-contact distance was $12\;$nm for the AFM experiment with a very stiff cantilever \cite{Zwol2008a}. It is much smaller than the minimal distance for high precision experiments  ( from $40\;$nm \cite{Torricelli2010} to $150\;$nm \cite{Decca2005}). The surface force apparatus \cite{Israelachvili2010} that uses much stiffer springs is more stable, but even in this case the shortest distance (in vacuum) was about $8.5\;$nm \cite{Tonck1991}. Only two measurements exist at these short separations between good conductors (Au-Au) \cite{Zwol2008a,Tonck1991}, while a similar short range measurement was also performed between a very flat nitrogen doped SiC and low roughness Au surfaces \cite{Sedighi2016}. A precision of the short-distance experiments is low because the absolute uncertainty in the separation $\delta d\sim 1\;$nm gives a large relative error in the force measurement $\sim \alpha(\Delta d/d)$, where $2<\alpha<3$ for the sphere-plate configuration. Due to the pull-in instability the data in the short-distance range $d\sim 10\;$nm are sparse and have poor precision, although in this range the role of the DF is crucial for many physical phenomena.

%============================================================

%===========================================================

To predict the force theoretically at short distances is as difficult as to measure it. The problem occurs due to non-additivity of the DF. "In condensed bodies (in contrast with gases) the atoms in the neighbourhood cause an essential change in the properties of the electronic shells, and the presence of a medium between the interacting atoms affects the electromagnetic field fluctuations through which the interaction is established" \cite{DLP}. While the root-mean-square (rms) roughness is small in comparison with the distance between the bodies, the roughness contribution can be evaluated as a perturbative effect \cite{Genet2003,MaiaNeto2005}. However, when the roughness amplitude is comparable with the separation gap, there is no a reliable way to predict the force (see review \cite{Svetovoy2015} on the non-additive roughness effects).

The non-perturbative roughness contribution was observed \cite{Zwol2008b} as a strong deviation from the expected power-law scaling for rough bodies close to contact. Qualitatively this effect was observed even earlier \cite{DelRio2005}. An approach to deal with the roughness problem theoretically was proposed by Broer \textit{et al.} \cite{Broer2011} and then developed in more detail \cite{Broer2012}. The idea of the method is based on a so called “grass and trees” model. High peaks (trees), which define the minimum distance between the bodies (distance upon contact), are rare and the average distance between them is large. For this reason their contribution can be calculated additively. On the other hand, the asperities with heights close to the rms roughness (grass) can be calculated using the perturbation theory. This approach successfully reproduced the experimental results for very rough gold surfaces \cite{Broer2012}, but the experimental precision was not sufficient to check the model for smoother surfaces, and discriminate between the additive and non-additive contributions. Moreover, roughness of deposited gold films is described by the extreme value statistics \cite{Zwol2009}, and it is not clear if one can apply the model for other statistics as well.

In this paper we propose a method to measure the dispersion forces between rough solids close to contact. The method is based on the adhered cantilever and does not suffer from the loss of stability at short separations. It allows also to test models for evaluation of the roughness contribution to the dispersion forces at distances comparable with the roughness amplitude.

\section{Method of adhered cantilever}

\begin{figure}[tb]
\begin{center}
%\vspace{0.3cm}
\includegraphics[width=0.7\textwidth]{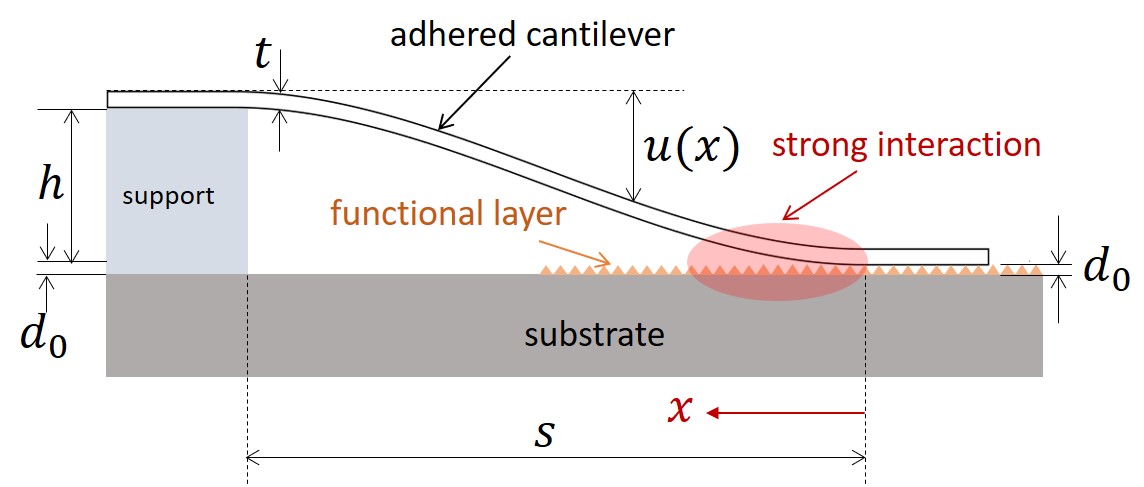}
\caption{Adhered cantilever. The main parameters and the choice of the coordinate system are shown. Domain, where the dispersion force between the cantilever and substrate is strong, is highlighted. The minimum distance $d_0$ between the cantilever and substrate is established by roughness of the solids. \label{fig:adhered}}
\end{center}
\end{figure}

The stiction problem in MEMS has been analysed experimentally and theoretically \cite{Maboudian1997,Parker2005} for a model system that is the adhered cantilever shown in Fig.~\ref{fig:adhered}. One end of a rectangular flexible beam is firmly fixed at a height $h$ above a substrate. If the beam is long enough, the other end will stick to the substrate after the last step of the fabrication process (drying). It was demonstrated \cite{Mastrangelo1993a,Mastrangelo1993b} that the adhesion energy per unit area $\Gamma$ is related to the unadhered part of the cantilever with the length $s$ and using this relation an accurate method to determine $\Gamma$ has been proposed \cite{Boer1999}. Simultaneous influence of the adhesion and electrostatic forces on the shape of the cantilever also has been addressed \cite{Knapp2002}.

%============================================================

%===========================================================

\subsection{Idea of the method}

In the adhered area the attractive forces (dispersion, electrostatic, capillary, chemical) are equilibrated by the repulsion due to elastic deformation of roughness asperities of the solids in contact. Outside of the adhered area but close to it (highlighted area in Fig.~\ref{fig:adhered}) the vdW/Casimir force between the beam and substrate is equilibrated by the elastic force  due to bending of the cantilever. This force has to influence the shape of the cantilever. Moreover, this influence has to be appreciable far away from the adhered end due to the boundary condition $du/dx=0$ at $x=0$, where $u(x)$ is the shape of the cantilever as shown in Fig.~\ref{fig:adhered}. Since the adhered cantilever is always in a stable state, one can measure the forces acting near the adhered end without the distance restrictions. It can be done analysing the shape of the cantilever. This is the main point of the proposed method, but to see its feasibility we have to consider the problem in more detail.

An attempt to calculate numerically the change of the shape due to vdW forces has been undertaken by Knapp and de Boer \cite{Knapp2002}. However, these authors did calculations in an unfavorable range of parameters and concluded that the effect is negligible. More detailed analytic and numerical analysis \cite{Svetovoy2017} demonstrated existence of the parameter range, in which the force influences the shape of the cantilever on a measurable level.

\subsection{Shape of the cantilever}

\begin{figure}[tb]
\begin{center}
%\vspace{0.3cm}
\includegraphics[width=0.7\textwidth]{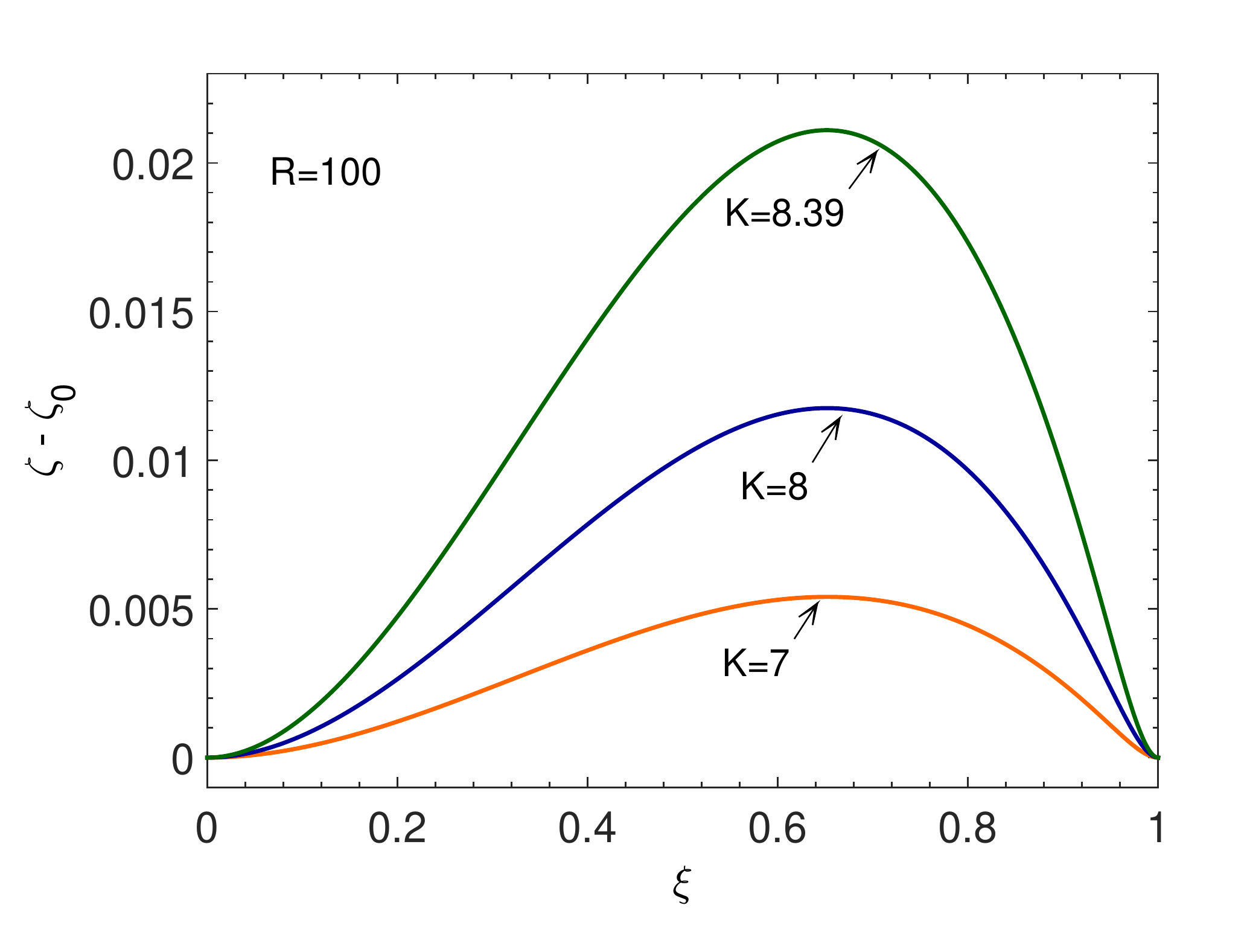}
\caption{Normalized deviation of the shape of the cantilever $\zeta=u(x)/h$ from the classic shape $\zeta_0=u_0(x)/h$ as a function of the lateral coordinate $\xi=x/s$. The results are presented for a few values of the parameter $K$, which is proportional to the dispersion pressure $P_0$ at the minimum separation $d_0$. The figure is taken from Ref. \cite{Svetovoy2017}.  \label{fig:deviation}}
\end{center}
\end{figure}

If we completely neglect external forces acting on the unadhered part, the cantilever has the classic shape
\begin{equation}\label{eq:classic}
 u_0(x)=h(1-3\xi^2+2\xi^3),\ \ \ \xi=x/h
\end{equation}
that is defined by the balance of the adhesion and elastic forces \cite{Knapp2002}. In reality outside of the adhered area the cantilever interacts with the substrate via the gas (or liquid) gap due to the DF and the residual electrostatic forces. The latter can appear as a contact potential difference or as surface charges due to trapped charge in dielectrics. The electrostatic force can be compensated as discussed below (see Sec.~\ref{compensation}) so that in the external force we include only the DF. Solving the equation of the elasticity theory one can find the correction to the classical shape due to the external force \cite{Svetovoy2017}. This correction is shown in Fig.~\ref{fig:deviation}. For simplicity the DF per unit area was presented in the form that is well justified in the restricted range of distances \cite{Palasantzas2010}
\begin{equation}\label{eq:force}
  P(d)=P_0(d/d_0)^{-\alpha},
\end{equation}
where for the plate-plate interaction the exponent $\alpha=3$ corresponds to pure vdW force and  $\alpha=4$ corresponds to pure Casimir force. The actual situation can be described by an intermediate parameter $3<\alpha<4$. The parameter $P_0$ has the meaning of the dispersion pressure at the minimum distance $d_0$. An important parameter characterising the problem is also $R=h/d_0\gg 1$, which is always large for any practical situation.

%============================================================

%===========================================================

As one can see in Fig.~\ref{fig:deviation} the deviation $\Delta u=u(x)-u_0(x)$ is maximum at $x=s/3$ that is far away from the adhered end. The width of the domain, where the force is strong, is estimated as the lateral size of the region where the vertical distance between the beam and substrate is $d<2d_0$ that corresponds the force reduction roughly one order of magnitude. It gives the width $\sim s\sqrt{d_0/h}\ll s/3$. This property is especially convenient for the force measurement. The relative value of $\Delta u$ at maximum reaches percents and strongly depends on the parameter $K$ that is defined as the ratio of the dispersion pressure at the smallest distance $d_0$ to an elastic pressure $P_s$
\begin{equation}\label{eq:K_def}
  K=(P_0/P_s)^{1/4},\ \ \ P_s=Et^3h/12s^4,
\end{equation}
where $E$ is Young's modulus of the beam material and $t$ is the thickness of the cantilever. When the parameter $K$ reaches a critical value $K_c$ the deviation $\Delta u$ becomes maximum and scales as $\Delta u\sim \sqrt{d_0 h}$. The value of $K>K_c$ cannot be reached because in response to a too large force the unadhered length $s$  will be reduced resulting in $K$ smaller than the critical value. Therefore, the experiment has to be designed to keep the value of $K$ as close as possible to $K_c$ where the sensitivity to the DF acting near the adhered end is the largest.

\subsection{Work done by the DF}

To determine the force one can measure the unadhered length $s$ and the maximum deviation from the classic shape $\Delta u_{max}=\Delta u(s/3)$. From these two values it is possible to find the force at the minimum distance $P_0$ and the adhesion energy $\Gamma$. Because bending of the cantilever depends on the force in an integral way, the deviation $\Delta u$ is defined only by the force at $d_0$. The adhesion energy $\Gamma$ depends weakly on the external force and we can use the classic relation \cite{Boer1999}
\begin{equation}\label{eq:Gamma_def}
  \Gamma\approx\frac{3Et^3h^2}{2s^4}.
\end{equation}
The relative correction to this relation due to the DF \cite{Svetovoy2017} scales as $R^{-3/2}$.

Combining Eqs.~(\ref{eq:K_def}) and (\ref{eq:Gamma_def}) one can present the parameter $K$ via $\Gamma$ and $P_0$ as
\begin{equation}\label{eq:K_via_GP}
  K^4=18R\left(\frac{d_0P_0}{\Gamma}\right).
\end{equation}
As follows from (\ref{eq:force}) the work done against the DF to put the cantilever at the position $d=d_0$ is equal
\begin{equation}\label{eq:work}
  W=d_0P_0/(\alpha-1).
\end{equation}
Therefore, $K^4$ is proportional to the ratio of $W$ to the adhesion energy. In general, there are a number of sources contributing to the adhesion energy, but, if the adhesion is defined only by the dispersion interaction, $\Gamma=W$ will get the smallest value resulting in the largest $K$ (highest sensitivity). For this case we find $K=2.59R^{1/4}$, while the critical value determined in \cite{Svetovoy2017}  is $K_c=2.65R^{1/4}$ (both values are presented for $\alpha=3.5$). We see that the highest sensitivity to the force is realized for the weakest adhesion, when $K$ is close to the critical value. The possibility to reach the weakest adhesion was demonstrated experimentally \cite{DelRio2005}. On the other hand, for strong adhesion $\Gamma\gg W$ the shape of the cantilever approaches the classic one and the method of the adhered cantilever will not be sensitive to the DF anymore.

\subsection{Advantages and disadvantages of the method}

The method of adhered cantilever has the following advantages over the method of elastic suspension.
\begin{enumerate}[(i)]
  \item No pull-in instability at short distances. It is the most important property that allows measurement of the forces at short distances, which are difficult or impossible to examine by the elastic suspention method.
  \item The force and the adhesion energy are measured simultaneously. This property discriminates contribution to the adhesion energy that differs from the dispersion force.
  \item The force is measured between practically parallel plates. No parallelism problem appears in contrast with the elastic suspension method.
  \item No restriction on the used materials. It can be metals, semiconductors, or dielectrics and any combination of two different materials. Due to trapped charges in dielectrics measurement of the DF is problematic for the elastic suspension.
  \item The force can be measured in gas or liquid environment with a comparable precision. The method of elastic suspension suffers from a significant loss of precision if the force is measured in a liquid.
\end{enumerate}
However, there are disadvantages of the adhered cantilever method.
\begin{enumerate}[(a)]
  \item The force can be measured only at one specific distance $d_0$ corresponding to the minimum distance between solids. This distance can be changed only by changing roughness of interacting solids.
  \item The method is sensitive to accidental nanoparticles in the adhered area, although the presence of these particles can be easily recognized. It demands assembling of the measuring chips in very clean conditions.
\end{enumerate}

\section{Configuration of the experiment}

\subsection{Optimal dimensions}

Since originally the adhered cantilevers have been used as a model system to analyse the stiction problem in MEMS, dimensions of the cantilevers were typical for micromechanics \cite{Maboudian1997,Knapp2002,DelRio2005}: the width $w=20-30\;\mu$m, thickness $t=2-3\;\mu$m, support height $h=2\;\mu$m, and total length $L=1000-1500\;\mu$m.  For $d_0=10\;$nm and $h=2\;\mu$m the maximum deviation from the classical shape is estimated as \cite{Svetovoy2017} $\Delta u_{max}\approx 34\;$nm (see Eq.~(\ref{fig:deviation}) below). This value is measurable by any interferometric method, but to increase precision and widen available range of parameters it is preferable to have larger deviations.

It can be done by increasing the support height $h$ to a value of $10-20\;\mu$m when $\Delta u_{max}=77-108\;$nm. Moreover, thin cantilevers made of polysilicon by microfabrication can deviate from ideal behavior due to a nonzero take-off angle at the fixed end, finite torsional compliance, and initial nonzero curvature of the unadhered cantilever \cite{Jensen2001}. This nonideality also can influence the shape of the cantilever introducing significant uncertainties. To reduce strongly these effects we propose to fabricate the cantilevers from a single crystal silicon. It is possible to do using silicon-on-insulator (SOI) wafers, for which the thickness of the top layer Si above SiO$_2$ is $t=10-15\;\mu$m. Using for an estimate $t=h=10\;\mu$m and a typical value of the adhesion energy induced by the dispersion forces $\Gamma=10\;\mu$J/m$^2$ one finds from Eq.~(\ref{eq:Gamma_def}) the unadhered lenth of the cantilever $s\approx 7\;$mm. From this estimate we can conclude that the total length of the cantilever has to be about $L=10\;$mm. A convenient width for these long cantilevers is $w=1\;$mm.
In comparison with microcantilevers suggested by MEMS these single crystal minibeams are sufficiently soft but more convenient for handling. Moreover, due to a large adhered area the adhesion energy $\Gamma$ is a well defined parameter with a small place-to-place variation.

\subsection{Chip design}\label{chip}

\begin{figure}[tb]
\begin{center}
%\vspace{0.3cm}
\includegraphics[width=0.7\textwidth]{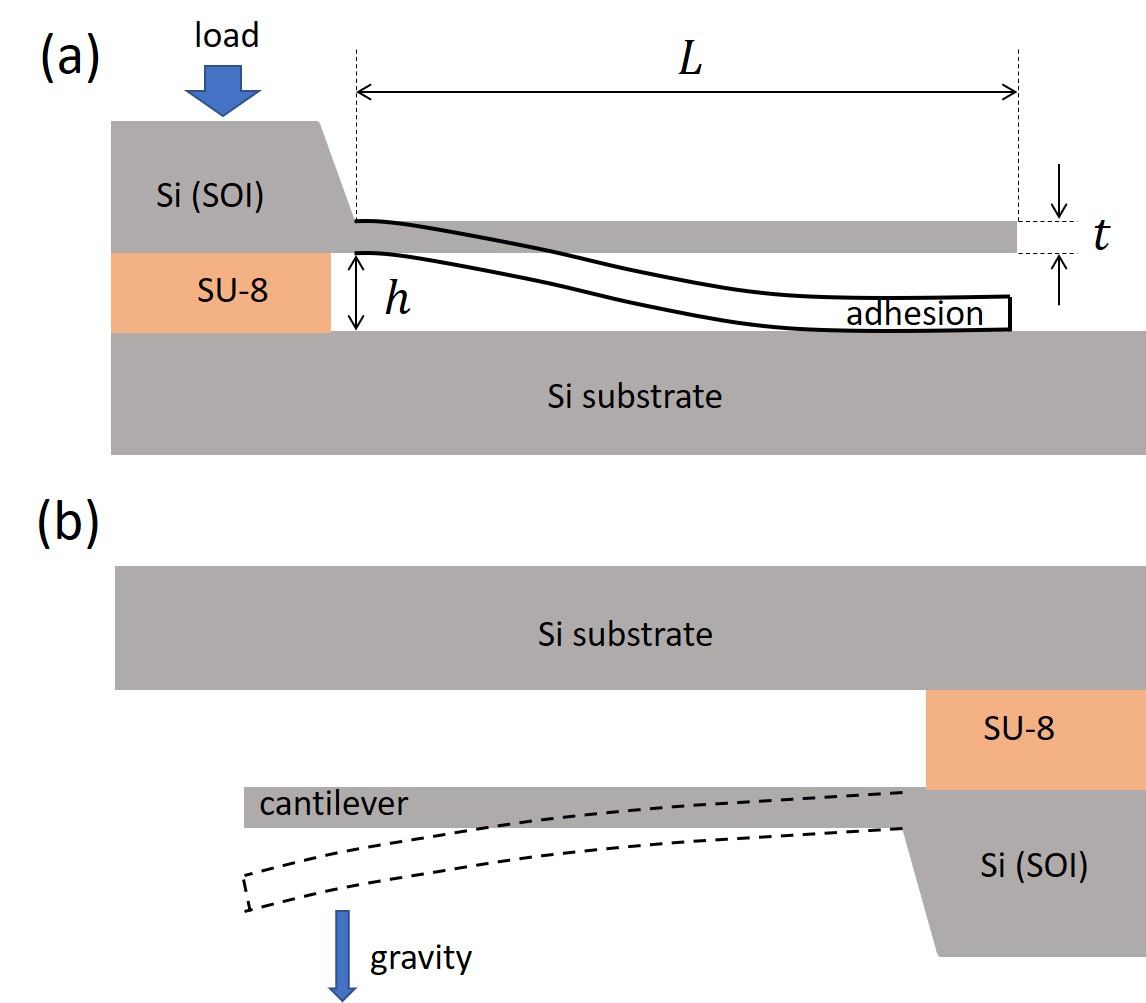}

\caption{(a) Measuring unit. The chip is assembled from SOI wafer containing the single crystal Si cantilever and a common Si wafer separated by a layer of SU-8. (b) Upside down assembling of the wafers to prevent initial adhesion.  \label{fig:assemble}}
\end{center}
\end{figure}

The chips for measurements will be assembled from two silicon wafers as shown in Fig.~\ref{fig:assemble}. The top layer is a polished Si with a thickness $t$ that is separated from the base Si layer by a thin silicon dioxide layer. The cantilevers are fabricated in the top layer and the base layer under the cantilevers is etched away by plasma etching using SiO$_2$ as a stop layer. The cantilevers are released by wet etching of SiO$_2$ in HF. The second wafer is an ordinary Si wafer covered with the negative  SU-8 photoresist of thickness $h$, which is used as the support for the cantilevers. Both wafers are bonded together as shown in Fig.~\ref{fig:assemble}(b). The front sides of the wafers are facing each other and the cantilevers are deflected away from the landing pads due to gravity (the free end is deflected about $40\;\mu$m). Alignment is easy due to rather large lateral sizes of the beams. Rotating the pile topside up as shown in (a) will result in the adhesion of cantilevers induced by gravity.

%============================================================

%===========================================================

\begin{figure}[tb]
\begin{center}
%\vspace{0.3cm}
\includegraphics[width=0.7\textwidth]{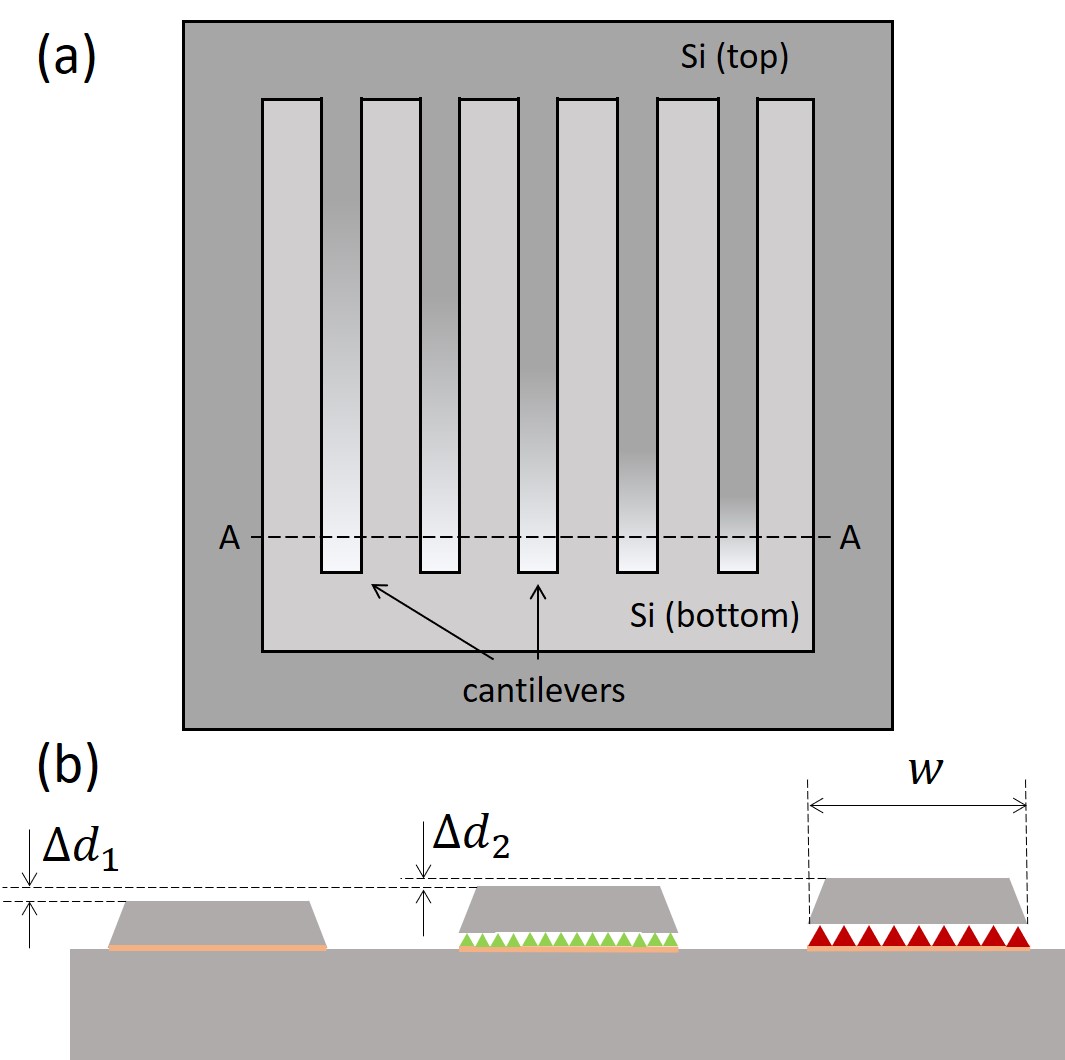}
\caption{(a) Top view of the chip. Each cantilever can have different unadhered length (shown by gradient of gray). Areas covered by the top and bottom wafers are indicated. (b) Cross-section along the AA line (adhered ends). Height difference is due to different functional layers at the landing pads.  \label{fig:chip}}
\end{center}
\end{figure}

One wafers of $100\;$mm in diameter will contain 12 chips with a size of $18\times18\;$mm. Each chip includes 5 identical beams as shown in Fig.~\ref{fig:chip}, but the ends of each beam and the corresponding landing pad can be covered with different functional layers. Due to different adhesion energy of the beam with the landing pad, the unadhered length $s$ for each cantilever in the series will vary. As functional layers we are planning to use different metals, semiconductors, dielectrics, and different kind of nanoparticles deposited on Si wafers. There are many possibilities to change roughness of the functional layer even for the same functional material. It gives the way to vary the minimum distance $d_0$ and, therefore, to measure the forces at different distances.

%============================================================

%===========================================================

\subsection{Characterization}

Before assembling the top and bottom wafers the ends of the cantilevers and the corresponding landing pads can be characterized for roughness with AFM. This possibility is an important advantage of the chip design. Since the minimum distance between the solids $d_0$ is defined by the highest asperities, one has to know not only the roughness distribution around the most probable height but also the tail of the distribution for the highest asperities. The latter is a nontrivial problem because to collect a reliable statistics for rare events (high peaks) one has to analyse as large area as possible.  Nevertheless, it is feasible as was demonstrated in Ref.~\cite{Zwol2009} where Au films were investigated using AFM scans with a size of $20\times20\;\mu$m$^2$ and with a resolution of $4096\times4096\;$pixels. It was found that high peaks on gold films are well described by the extreme value statistics rather than by the tail of the normal distribution. For the adhered cantilever experiment there is room to improve the data collection and transformation.

Furthermore, it is important to know with the best possible precision the minimum distance between the solids $d_0$. It is essential because uncertainties in $d_0$ define the relative error in the force measurement. It can be done by scanning with the laser beam of an interferometer along the line AA as shown in Fig.~\ref{fig:chip}(b). Because the total thickness to measure is about $10\;\mu$m but the target precision is $1\;$nm we need the interferometer with a high dynamic range. For this purpose one can use a homodyne Michelson interferometer with quadrature signals \cite{Reibold1981,Pozar2011}. Feasibility of this method was checked with a homemade interferometer \cite{Uvarov2018} of this kind (see Sec.~\ref{shape}). The same interferometer can be used for determination eigen frequencies and quality factors of the cantilevers recording free oscillations in air or/and in vacuum excited with a piezo. These values will provide auxiliary information on the cantilevers.

Among five cantilevers in the chip one can be used to calibrate the distance $d_0$. This cantilever and the corresponding landing pad have no any functional layers. Therefore, in the adhered area two very smooth (rms roughness $0.2-0.3\;$nm) Si surfaces meet. The average separation distance between these surfaces can be reliably predicted using proximity force approximation (PFA) and detailed information on the surface roughness. The values of $d_0$ for all the other cantilevers on the chip can be determined from the interferometric measurements as shown in Fig.~\ref{fig:chip}(b).

An additional important characteristic of the cantilever is the uniformity of the thickness $t$. For SOI wafers the produces guarantee the thickness of the top working layer at the level of $\pm 0.5\:\mu$m over the entire wafer scale. It is much larger than the deviations from the classical shape, which we are going to measure. Therefore, the thickness of the cantilevers has to be carefully characterized over their length. It can be done using the wavelength $\lambda=1.15\:\mu$m (HeNe laser), for which silicon is transparant. Interference of the light reflected from the top and from the bottom sides of the cantilever carries information on the thickness $t$ at a given position of the laser beam.

\subsection{Shape of the cantilever}\label{shape}

The cantilever beam has the classical shape (\ref{eq:classic}) if the adhesion energy $\Gamma$ is large in comparison with the work $W$ (\ref{eq:work}) done by the DF. The classic shape can be tested for strong adhesion induced, for example, by capillary interaction. To this end hydrophilic cantilevers can be tested in humid atmosphere. To keep the unadhered length $s$ in the same range, one can increase the thickness $h$ of the separating SU-8 layer for some chips. The effects from nonzero take-off angle and finite torsional compliance \cite{Jensen2001} have to be small because the cantilevers are made from a single piece of crystalline silicon, but even small effects can be characterized with the interferometer and taken into account in the analysis. The same can be done with the initial curvature of the cantilever that is \textit{a priory} is not negligible.

\begin{figure}[tb]
\begin{center}
%\vspace{0.3cm}
\includegraphics[width=0.7\textwidth]{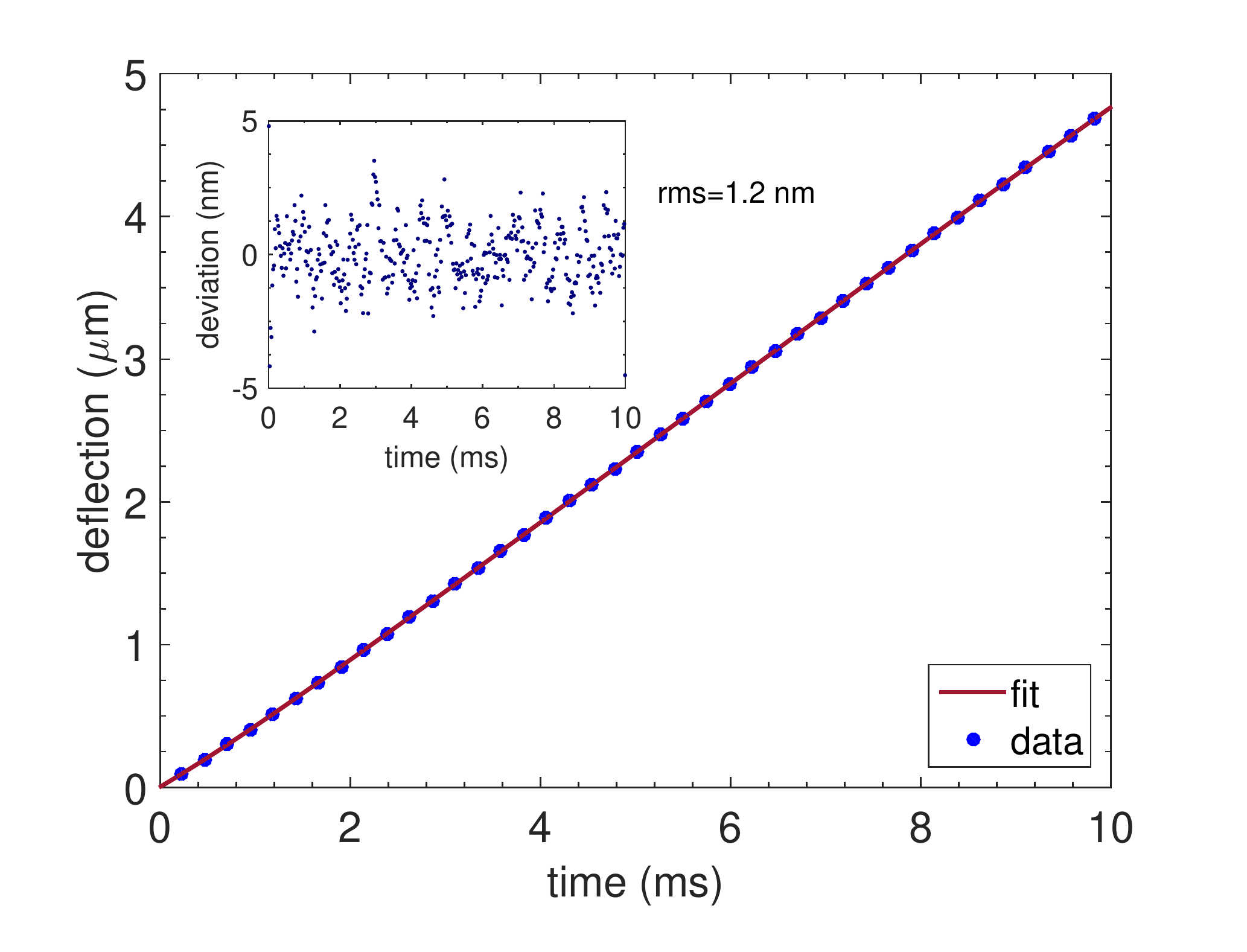}
\caption{Deflection of the actuator membrane \cite{Uvarov2018} with time measured by the quadrature signal interferometer as a function of time. The best fit of the data is shown by the solid line. Only 10\% of the data are shown. Deviation between the data (all points) and the fit is shown in the inset. The rms deviation is only $1.2\;$nm while the total deflection is $4.7\;\mu$m. This figure is original although it uses the data from \cite{Uvarov2018}. \label{fig:membrane}}
\end{center}
\end{figure}

The shape of an adhered beam can be measured by scanning along the cantilever with the laser beam of the interferometer. The quadrature signal interferometer is able to measure the change in the optical path with a precision of $1\;$nm in a wide dynamic range. The instrument has been used previously \cite{Uvarov2018} to observe the movement of a flexible membrane driven by the electrochemical process. Figure~\ref{fig:membrane} shows the data for the deflection of the membrane with time and the best fit for this nearly linear process. The difference of the data and the fit is shown in the inset. One can see that the rms noise stays within $1.2\;$nm while the membrane deflects from zero to $4.7\;\mu$m. It corresponds to a dynamic range of the interferometer of about $70\;$dB, but it can be increased further to 80 dB. This is because a small deflecting membrane is less convenient for observation than a flat cantilever that bends on a very small angle. This example demonstrates feasibility to measure the bending of cantilever up to $10\;\mu$m with a precision of  $1\;$nm.

%============================================================

%===========================================================

From the scan of a specific cantilever one can determine the undhered length $s$. Comparing the scan with the expected classic shape corrected to the initial nonzero curvature and nonhomogeneous thickness, one finds the maximum deviation from the classic shape at $x=s/3$. From the theory this deviation is expected as \cite{Svetovoy2017}
\begin{equation}\label{eq:deviation}
  \Delta u_{max}\approx (2/3)^{7/2}\sqrt{d_0h}\times\frac{d_0P_0}{\Gamma (\alpha-1)}.
\end{equation}
We see from this relation that the largest deviation is possible for the smallest $\Gamma$. The adhesion energy is the smallest for the case when it is defined completely by the dispersion forces $\Gamma=W$. Using Eq.~(\ref{eq:work}) we find that the last factor is equal to one. The method of adhered cantilever is favorable while $\Gamma \sim W$, but in the limit $\Gamma \gg W$ when strong adhesion forces dominate (for example, capillary) the condition for determination $P_0$ becomes unfavorable.

Let us estimate the expected values of $W$ for Si-Si and Au-Au interaction at the distance $d_0=10\;$nm. Assuming pure vdW interaction $\alpha=3$ (true only as a rough estimate) one has $W=A_H/12\pi d_0^2$, where $A_H$ is the Hamaker constant for interacting materials. For Si-Si this constant is \cite{Parsegian2006} $A_H=2.6\times10^{-19}\;$J and for Au-Au it is \cite{Klimchitskaya2000} $A_H=4.5\times10^{-19}\;$J. Then for the interaction energy between ideally flat surfaces one finds $W=69\;\mu$J/m$^2$ for Si-Si and $W=119\;\mu$J/m$^2$ for Au-Au interaction. The unadhered length $s$ increases with the thickness of cantilever $t$, while this length decreases with the increase of the  adhesion energy. Thus, we can always choose a proper cantilever thickness $t$ (or support height $h$) to work in the desired range of $W$.

From Eq.~(\ref{eq:deviation}) the dispersion pressure $P_0$ at the minimum distance is expressed via the directly measurable parameters $s$, $\Delta u_{max}$, and $d_0$. Introducing uncertainties for each of the parameter by symbol $\delta$ one can express the relative error in the dispersion pressure as
\begin{equation}\label{eq:error}
  \frac{\delta P_0}{P_0}=\left[\left(\frac{4\delta s}{s}\right)^2+\left(\frac{\delta u}{\Delta u_{max}}\right)^2+\left(\frac{3\delta d_0}{2d_0}\right)^2\right]^{1/2}.
\end{equation}
The relative error uncertainty in the unadhered length $s$ is estimated as 1\%. The shape of the cantilever can be measured with a precision of $\delta u=1\;$nm. If the adhesion energy $\Gamma\sim W$, then the relative error in the shape is about $2$\%. The error in $d_0$ is the most important. It is defined by the precision, with which we can measure the difference in the heights of the functional layers, that is $\delta d_0=1\;$nm. This is a typical value realized for all the Casimir force experiments \cite{Harris2000,Chan2001,Decca2003,Decca2005,Iannuzzi2004,Chen2007,Man2009,Torricelli2010}. Since in our case $d_0$ is rather small the relative error in the pressure due to $d_0$ is estimated as 15\%. It has to be stressed that using the elastic suspension method the same error in the pressure would be $\alpha\delta d_0/d_0$ that is more then twice larger ($\alpha>3$). It occurs because the adhered cantilever method is sensitive to the integral effect of the force but not directly to the force at a given distance.

\subsection{Alternative determination of $d_0$}

The force measured even with a poor precision at such short distances is of a considerable interest. However, we propose a way to reduce $\delta d_0$ that is based on a large nominal area of contact for our cantilevers. For this purpose we can combine the detailed roughness statistics of the contacting bodies with the prediction relied  on the contact mechanics. The basis for this approach was proposed \cite{Zwol2008a}  and developed earlier \cite{Zwol2009} but only for a negligible load. Finite adhesion energy provides a significant load on high asperities, which get into contact with the opposing solid. These asperities are deformed elastically and possibly plastically and their number can be determined from the roughness statistics.

%Contact of two rough plates is equivalent to the contact of a flat hard plate and an elastic rough plate with a combined roughness topography and effective Young’s modulus.

\begin{figure}[tb]
\begin{center}
%\vspace{0.3cm}
\includegraphics[width=0.7\textwidth]{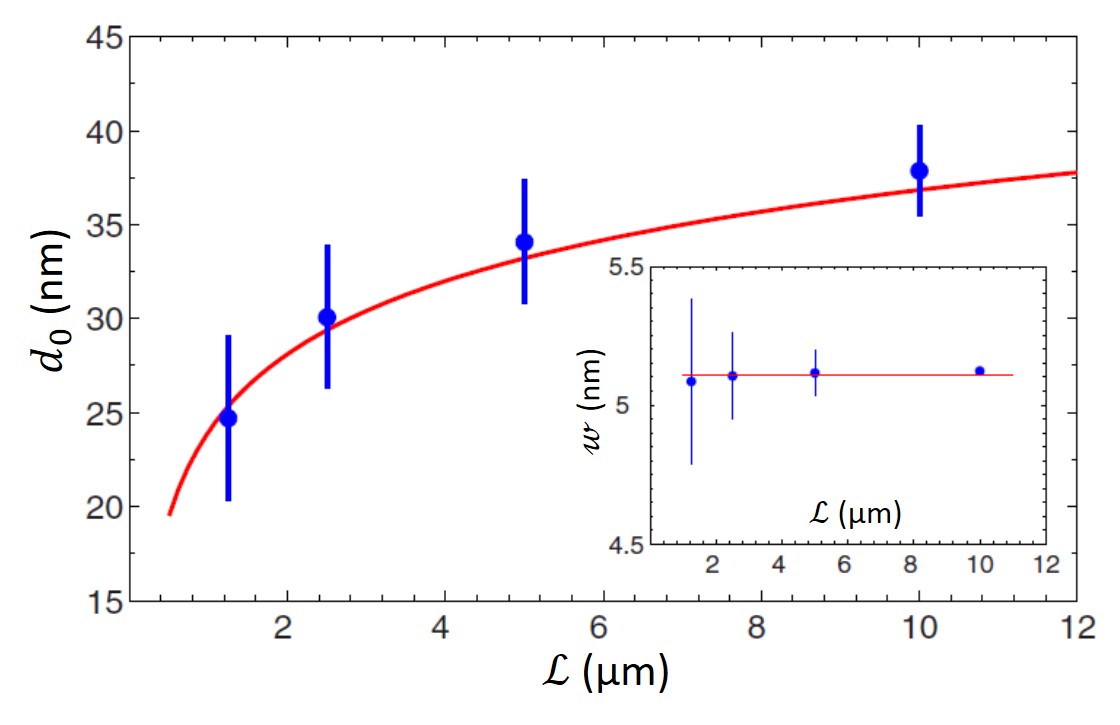}
\caption{Minimum distance $d_0$ between two plates  allowed by their roughness at zero load. This distance depends on the analyzed area size $\cal{L}$  and varies from place-to-place (blue dots with error bars). The data were taken from AFM megascans  of two gold films. The red curve is a theoretical prediction. The inset shows that the rms roughness does not depend on the size $\cal{L}$. The figure is taken from Ref.~\cite{Zwol2009}. \label{fig:contact}}
\end{center}
\end{figure}

A rough plate is described as a set of asperities with a random height and with a lateral size that is given by the correlation length determined from the roughness topography. The height distribution is determined from the roughness analysed over as large area as possible. Large area is needed to collect reliable statistics of rare events, namely high asperities. This statistics can differ significantly from the normal distribution as was demonstrated for deposited gold films  \cite{Zwol2009}. It is important that the number of high asperities, which can be in contact with the opposing surface, depends on the analysed area. Figure~\ref{fig:contact} shows the height of the highest asperity (equivalent to $d_0$ in the limit of zero load) as a function of the area size $\cal{L}$. One can see that this height increases with the size and has a significant rms deviation due to place-to-place variation of the highest peak. It has to be stressed  that the rms deviation of $d_0$ decreases with the size $\cal{L}$. For the nominal adhered area $\sim 1\;$mm$^2$, which will be used in the proposed experiment, we expect variation of $d_0$ considerably smaller than $1\;$nm.

%============================================================

%===========================================================

The adhesion energy, which is extracted from the unadhered length of the cantilever, has to be equal to the free energy of $N$ elastically deformed asperities with the height $d_0$. All peaks that are higher than $d_0$ are deformed plastically and their height is reduced to $d_0$. The number $N$ of the asperities with the height $z>d_0$ within the area $\cal{L}\times\cal{L}$ is defined as \cite{Zwol2009} $N=({\cal{L}}^2/l_c^2)(1-{\cal{P}}(d_0))$, where $l_c$ is the correlation length and ${\cal{P}}(z)$ is the cumulative distribution of the asperity heights. The relative elastic deformation of these $N$ asperities is determined from the balance of the elastic and adhesion energies. The same relative deformation defines the pressure acting on each asperity. In the simplest approach, when a peak is approximated by a bar with the cross section area $l_c^2$, this pressure $p$ is expressed as
\begin{equation}\label{eq:bar}
  p=\sqrt{(2\Gamma /d_0)(1-{\cal{P}}(d_0 ))}.
\end{equation}
It has to be equal to the plastic yield strength to prevent further deformation of the asperities. Equation (\ref{eq:bar}) can be considered as the equation for the minimum distance $d_0$. Even this simplified model predicts reasonable values of $d_0$. For example, using the cumulative probability for $800\;$nm thick gold film (see \cite{Zwol2009,Broer2012}) and the adhesion energy induced by the dispersion interaction we find from (\ref{eq:bar}) $d_0=33.7\;$nm, while the value determined from the electrostatic calibration \cite{Zwol2008b} for this film is $d_0=34.5\pm1.7\;$nm.

This simple approach can be elaborated in detail. Realistic stress-strain curves can be used for each material; asperities can be modelled by hills of conical shape with rounded tips. Our aim is to reduce the error for determination of $d_0$ to a value of $\delta d_0\cong 0.2-0.3\;$nm that is given by the roughness of Si wafers. In this case the dispersion force at $d_0=10\;$nm can be determined with a relative precision of ~5\%. Prediction of $d_0$ based on the roughness topography and contact mechanics can be checked using the interferometric method for rather rough films. If the roughness is high, $\delta d_0\sim 1\;$nm determined interferometrically will be much smaller than $d_0$. Therefore, one can compare the prediction based on the topography with the interferometric measurement.

\subsection{Compensation of electrostatic contribution}\label{compensation}

The residual electrostatic force is the most important background force that has to be compensated in all experiments measuring the Casimir force. This residual force appears due to finite contact potential difference even between identical metals due to difference in wiring. The potential difference can be as high as a few hundred millivolts \cite{Lamoreaux1997,Decca2003} and is reduced to a few tens of millivolts in the best cases \cite{Decca2005}. The charges trapped in dielectrics made impossible the Casimir force measurement involving dielectric materials. Doping of dielectrics is used as a way to resolve the problem \cite{Sedighi2016}. At short separations the electrostatic effect is less severe because the DF is much stronger, but compensation of the electrostatic force has to be done in this case too.

\begin{figure}[tb]
\begin{center}
%\vspace{0.3cm}
\includegraphics[width=0.7\textwidth]{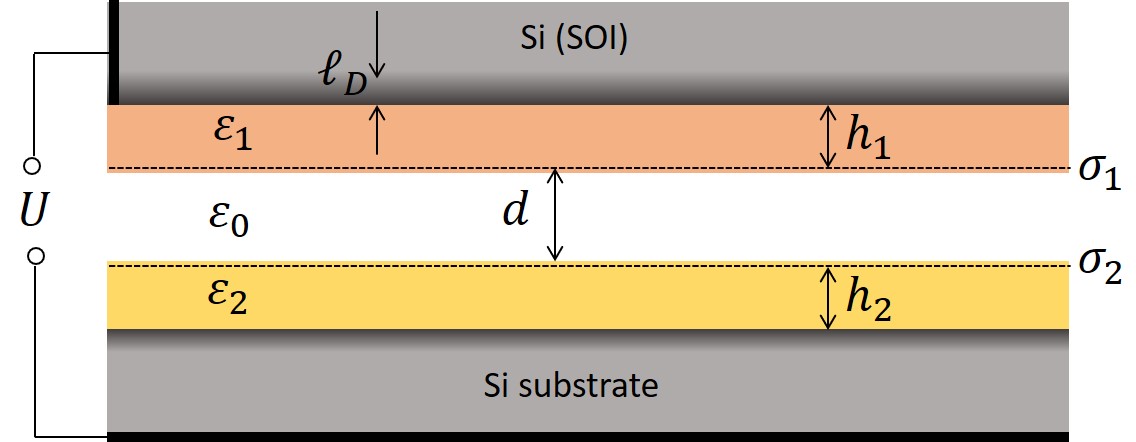}
\caption{Formation of the electrostatic pressure between solids approaching the contact. The Debye layer in Si is shown by the gradient of gray. The dashed lines show the layers of surface charges with densities $\sigma_{1,2}$. External potential difference $U$ is applied between Si wafers. \label{fig:electrostatic}}
\end{center}
\end{figure}

For the chip discussed in Sec.~\ref{chip} one can make the electrical contacts to the layer with cantilevers and to the bottom of the chip. Silicon has a finite conductivity due to doping and is characterized by a $13\;$nm thick Debye layer $l_D$ for a carrier concentration of $10^{17}\;$cm$^{-3}$. If the functional layers on the cantilever and landing pad both are dielectric, they can be charged with the surface density $\sigma_1$ and $\sigma_2$, respectively, as shown in Fig.~\ref{fig:electrostatic}. Neglecting for simplicity the Debye layers in Si, the electric field in the gap $d$ can be presented as
\begin{equation}\label{eq:field}
  E=-\frac{U+\frac{4\pi\sigma_1}{\varepsilon_1}h_1-\frac{4\pi\sigma_2}{\varepsilon_2}h_2}
  {d+\frac{\varepsilon_0}{\varepsilon_1}h_1+\frac{\varepsilon_0}{\varepsilon_2}h_2},
\end{equation}
where $\varepsilon_{1,2}$ and $h_{1,2}$ are the dielectric constants and thicknesses of the dielectrics and $\varepsilon_0$ is the permittivity of the intervening gas or liquid. The electrostatic pressure produced by this field is $P_{el}=\varepsilon_0 E^2/8\pi$.

%============================================================

%===========================================================

Typical thickness of the deposited layers is $h_{1,2}\sim 100\;$nm and typical concentration of the surface charges is $n_s\sim 10^{12}\;$cm$^{-2}$. Without compensation ($U=0$) the field in the gap (for the case $h_2=0$ and $\sigma_2=0$) is estimated to be $E=4\pi\sigma_1/(\varepsilon_1d+\varepsilon_0 h_1) \sim10^8\;$V/m, where we took $\varepsilon_1=3$ and $\varepsilon_0=1$. This field is well above the electrical breakdown of air  $E_b\approx 3\times 10^6\;$V/m or even of vacuum $E_b\sim 10^7\;$V/m. It means that there will be exchange of charges between bodies tending to reduce the field up to $E_b$. For example, ions produced in air gap by the field will be adsorbed on the dielectric surface and screen the trapped charges. Thus, the resulting potential to compensate will be $E_b(\varepsilon_1d+\varepsilon_0 h)/ \varepsilon_1$, and it is estimated from one to a few hundreds of millivolts. It is interesting also to see the effect of the charging on the force. The uncompensated potential results in the electrostatic pressure $P_{el}=\varepsilon_0 E_b^2/8\pi\sim 10^3\;$Pa, while  the vdW pressure between Si surfaces at $d_0=10\;$nm is $P_{vdW}=1.4\times 10^4\;$Pa. Thus, the charges trapped in the dielectric generate the electrostatic force that is smaller but not negligible in comparison with the DF.

If metal layers are deposited on Si wafers, then the external potential has to compensate the difference between the metal work functions. This difference is typically a few hundreds millivolts up to $1\;$V and it is established by exchange of the carriers. One can do the compensation in the same way as at the Casimir force measurements: find a potential that minimizes the total force. In the case of the adhered cantilever the total force will be minimum if the unadhered length is maximum.

\section{Conclusions}

We considered the proposition to measure the dispersion forces (DF) at short separations using the method of adhered cantilever. The distances between rough solids close to contact is a very inconvenient to measure the forces by standard method of elastic suspension due to pull-in instability of the system. However, this distance range ($\sim 10\;$nm) is responsible for many physical phenomena because the DF are strong enough to compete with the electrostatic forces. The adhered cantilever can be used for measurements in this range of distances because it does not suffer from the instability. We explained the general idea that is based on sensitivity of the shape of the cantilever to the forces acting near the adhered end. The advantages of the method are: no loss of stability at short distances, the force and adhesion energy are measured simultaneously, the force is measured between practically parallel plates, no restrictions on the used materials, and the force in liquid can be measured with a similar precision as in gas.

To demonstrate the feasibility of the method we defined the optimal dimensions of the cantilever that guarantee the highest sensitivity to the DF. It was also stressed that the method is not applicable for strong adhesion between the solids induced by capillary or chemical interaction. It was argued that minicantilevers fabricated by micromachining from a piece of single crystalline silicon are able to provide precise measurement. The design of the measuring chip was developed.

The method measures directly the length of the unadhered part of the cantilever $s$, the maximum deflection from the classic shape of the cantilever $\Delta u_{max}$, and the minimum distance between the solids in contact $d_0$. All three values can be determined with the demanded precision by a Michelson interferometer with quadrature signals, which provides a necessary wide dynamic range. It was concluded that the most important error in the force follows from the uncertainty in $d_0$.

Finally, we discussed the possibility to reduce the uncertainty in $d_0$ based on the measured surface roughness and on the contact mechanics that accounts for deformation of contacting asperities. This method can be applied only for rather large adhered area $\sim 1\;$mm$^2$ corresponding to the proposed size of the cantilevers. The residual electrostatic interaction is the main background. Although at short separations it is not as severe as at distances $d\sim 100\;$nm, it has to be compensated. The most difficult case is realized for dielectrics with trapped charges, but we proposed the way to compensate electrostatics including the case of trapped charges.
%\vspace{0.5cm}
\begin{acknowledgments}
This work is supported by the Russian Science Foundation, grant No.20-19-00214. G.~ P. acknowledges support from the Netherlands Organization for Scientific Research (NWO) under grant number 16PR3236.
\end{acknowledgments}

%\bibliography{Cantilever_lib}

\end{document}